\documentclass{article}
\usepackage{cite}
\usepackage{graphicx}
\usepackage{dcolumn}
\usepackage{caption}

\begin{document}

\date{}
\title{Allowed permutation symmetry in atomic and molecular wavefunctions. Simple
examples}
\author{Francisco M.
Fern\'andez\thanks{Francisco M. Fern\'andez,
fernande@quimica.unlp.edu.ar}\\ INIFTA, Divisi\'on Qu\'imica
Te\'orica\\Sucursal 4, Casilla de Correo 16\\1900 La Plata,
Argentina} \maketitle
\begin{abstract}
It is well known that the allowed wavefunctions for an
$N$-electron system should be antisymmetric with respect to the
permutation of any pair of electron labels. On the other hand, the
Hamiltonian for such system is invariant under any permutation of
electron labels and, consequently, its eigenfunctions are basis
for the irreducible representations of the symmetric group $S_N$.
Here, we investigate which symmetry species of the $S_N$ group are
compatible with the antisymmetry principle. We illustrate the
conclusions by means of simple $N$-particle one-dimensional models
with harmonic interactions.
\end{abstract}
\section{Introduction}

\label{sec:intro}

There has recently been a controversy about the permutation symmetry of
atomic and molecular Hamiltonians and the approaches commonly used to obtain
their eigenvalues and eigenfunctions. On one hand it has been stated that
Hartree-Fock and related methods do not take into account the permutation
symmetry of the non-relativistic Hamiltonian, which ``leads to false
concepts, misinterpretations and unjustifiable approximations when dealing
with many-electron systems''\cite{N19}. On the other hand, it has been shown
that the arguments put forward in that paper violate well known mathematical
theorems and that the analysis is based on an incorrect application of the
permutation operators of the symmetric group $S_{N}$\cite{F19}. Such
interesting discussion motivated the analysis of the permutation symmetry of
electronic systems carried out below in this paper.

The postulates of quantum mechanics state that the wavefunction for a system
of particles should be symmetric or antisymmetric under the permutation of
the variables of identical particles if they are either bosons or fermions,
respectively. In the particular case of an $N$-electron system the
wavefunction should be antisymmetric with respect to the transposition of
the coordinates of any pair of electrons. For this reason approximate
calculations of the electronic structure of atoms and molecules is commonly
based on Slater determinants constructed from suitably chosen spin-orbitals.
The configuration interaction (CI) method is known to provide accurate
atomic and molecular electronic energies\cite{P68}. On the other hand, it is
well known that the non-relativistic Hamiltonian for a system of $N$
electrons is invariant under the $N!$ permutations of the electronic
variables. For this reason its eigenfunctions are basis for the irreducible
representations (irreps) of the symmetric group $S_{N}$. Since the
Schr\"{o}dinger equation for an $N$-electron system is not exactly solvable
for $N>1$ there are no available comparisons between the exact solutions of
the non-relativistic system and sufficiently accurate results provided by
widely used methods like CI, except for some exactly-solvable models\cite
{M96}. It would be interesting, for example, to know to which
non-relativistic energy levels converges a CI calculation based on a
Slater-determinant basis set.

The purpose of this paper is to fill this gap by means of
exactly-solvable non-relativistic models with $S_{N}$ symmetry
that can be easily treated by means of CI to a great degree of
accuracy. In section~\ref{sec:perm_sym} we outline the concepts of
permutation symmetry that are relevant for present discussion. In
section~\ref{sec:N=3} we solve the Schr\"{o}dinger equation for a
simple non-relativistic model with $N=3$ identical particles. In
section~\ref{sec:N=4} we carry out a similar analysis for $N=4$.
Finally, in section~\ref{sec:conclusions} we summarize the main
results and draw conclusions.

\section{Permutation symmetry}

\label{sec:perm_sym}

The Hamiltonian operator $H$ for an $N$-electron system is invariant under
the transposition $P_{ij}$ of the variables of any pair of electrons $i$,$j$%
; that is to say: $P_{ij}HP_{ij}^{-1}=H$. There are $N(N-1)/2$ such
transpositions that satisfy $P_{ij}=P_{ji}=P_{ij}^{-1}$. Since $P_{ij}^{2}=%
\hat{E}$ (the identity operator) then the eigenvalues of every transposition
operator are $\pm 1$. The invariance of $H$ under transpositions can also be
written in terms of vanishing commutators $[H,P_{ij}]=0$. Since the
transpositions do not commute, then we cannot obtain a complete set of
eigenfunctions common to $H$ and all $P_{ij}$. We can write a transposition
as
\begin{equation}
P_{ij}=\left[
\begin{array}{l}
1,2,\ldots ,i,\ldots ,j,\ldots N \\
1,2,\ldots ,j,\ldots i,\ldots N
\end{array}
\right]
\end{equation}
which means to substitute the electron variables $\mathbf{r}_{j},\mathbf{r}%
_{i}$ for $\mathbf{r}_{i},\mathbf{r}_{j}$ (it may also include spin
variables when necessary)

The Hamiltonian $H$ is also invariant under any permutation
\begin{eqnarray}
P_{[i]} &=&\left[
\begin{array}{l}
1,2,\ldots ,N \\
i_{1},i_{2},\ldots ,i_{N}
\end{array}
\right]  \nonumber \\
i_{k} &\in &\{1,2,\ldots ,N\}
\end{eqnarray}
which means to substitute $\mathbf{r}_{i_{1}}$, $\mathbf{r}_{i_{2}}$,..., $%
\mathbf{r}_{i_{N}}$ for $\mathbf{r}_{1}$, $\mathbf{r}_{2}$,...,$\mathbf{r}%
_{N}$. There are $N!$ such permutations of the variables of the $N$
electrons that can be split into $N!/2$ even and $N!/2$ odd permutations.
Any permutation can be written as a non-unique product of a finite number of
transpositions\cite{CTDL77}. However, given a permutation, the number of
such factors is either even or odd and we commonly say that the permutation
is even or odd, respectively. The set of all $N!$ permutations of the $N$
electrons form the symmetric group $S_{N}$. The invariance of the
Hamiltonian may be expressed either as $P_{[i]}HP_{[i]}^{-1}=H$ or $\left[
H,P_{[i]}\right] =0$ for any of the $N!$ permutations.

If $\psi $ is an eigenfunction of $H$ with eigenvalue $E$, then $P_{ij}H\psi
=HP_{ij}\psi =EP_{ij}\psi $. Therefore, if $\psi $ is non-degenerate then $%
P_{ij}\psi =\pm \psi $ for all $i$, $j$. In the case of a degenerate energy
level both $\psi $ and $P_{ij}\psi $ may be linearly independent
eigenfunctions of $H$. In fact, since $\left[ P_{ij},P_{kl}\right] \neq 0$
then the non-degenerate states are not, in general, eigenfunctions of all
the permutation operators $P_{[i]}$. Despite of this fact it has been stated
that $\left[ H,P_{[i]}\right] =0$ implies that any eigenfunction of $H$ is
an eigenfunction of $P_{[i]}$\cite{N19,CBA03,BCA05}.

The Hamiltonians of some systems of identical particles are also invariant
under coordinate inversion $\hat{\imath}f(\mathbf{x})=f(-\mathbf{x})$ about
the origin. Since $[H,\hat{\imath}]=0$ and $[P_{ij},\hat{\imath}]=0$ then
the eigenfunctions of $H$ are either even or odd with respect to inversion: $%
\hat{\imath}\psi =\pm \psi $.

The results above apply to any system of $N$ identical particles but we
restrict ourselves to electrons because we are interested in the electronic
structure of atoms (with the nucleus clamped at origin) and molecules (under
the Born-Oppenheimer approximation). Since the Schr\"{o}dinger equation for
such systems cannot be solved exactly we resort to approximate methods. In
order to obtain a suitable basis set for such calculations we commonly
construct the required antisymmetric functions as Slater determinants\cite
{P68}
\begin{eqnarray}
\left| \chi _{i_{1}}\chi _{i_{2}}\ldots \chi _{i_{N}}\right\rangle &=&%
\mathcal{A}\chi _{i_{1}}(1)\chi _{i_{2}}(2)\ldots \chi _{i_{N}}(N)  \nonumber
\\
&=&\frac{1}{\sqrt{N!}}\sum_{i=1}^{N!}(-1)^{p_{[i]}}P_{[i]}\chi
_{i_{1}}(1)\chi _{i_{2}}(2)\ldots \chi _{i_{N}}(N)
\end{eqnarray}
where $p_{[i]}$ reflects the parity (even or odd) of $P_{[i]}$ and $\chi
_{j} $ is a spin-orbital given by the product of a space orbital factor $%
\varphi _{i}$ and a spin one $\omega _{k}$ that equals either $\alpha $ ($%
m_{s}=1/2$) or $\beta $ ($m_{s}=-1/2$). The CI method is a Rayleigh-Ritz
variational approach with the ansatz
\begin{equation}
\Phi =\sum_{i_{1},i_{2},\ldots ,i_{N}}C_{i_{1},i_{2},\ldots ,i_{N}}\left|
\chi _{i_{1}}\chi _{i_{2}}\ldots \chi _{i_{N}}\right\rangle  \label{eq:CI}
\end{equation}
commonly chosen to be an eigenfunction of the total spin operators $S^{2}$
and $S_{z}$ when $H$ is the non-relativistic (spin-free) Hamiltonian\cite
{P68}.

It is not possible to compare the approximate variational
calculation based on the trial function (\ref{eq:CI}) with an
exact result because the Schr\"{o}dinger equation for any atomic
or molecular system with $N>1$ cannot be solved exactly. For this
reason in the following sections we consider two exactly solvable
models with $S_N$ symmetry.

\section{Exactly-solvable three-particle model}

\label{sec:N=3}

The case $N=2$ is trivial because the only permutation operators are $\hat{E}
$ and $P_{12}$\cite{F19,CTDL77}. Therefore, any eigenfunction $\psi $ of the
non-relativistic Hamiltonian $H$ satisfies $P_{12}\psi =\pm \psi $ and a
symmetric spatial function is multiplied by an antisymmetric spin one
(singlet state), whereas an antisymmetric spatial function is multiplied by
a symmetric spin one (triplet) in order to obtain an antisymmetric total
wavefunction $\Phi $. Obviously, in this particular case we can easily omit
the spin part in the construction of an approximate wavefunction. Therefore,
the first non-trivial case is $N=3$.

The symmetric group $S_{3}$ is isomorphic to $C_{3v}$ (and also to $D_{3}$)%
\cite{C90}; its character table being
\[
\begin{array}{c|ccc|}
C_{3v} & \hat{E} & 2C_{3} & 3\sigma _{v} \\ \hline
A_{1} & 1 & 1 & 1 \\
A_{2} & 1 & 1 & -1 \\
E & 2 & -1 & 0
\end{array}
\]
There is a one-to-one correspondence between the permutation operators and
the $C_{3v}$ ones given by
\begin{eqnarray*}
\hat{E} &=&\left[
\begin{array}{l}
123 \\
123
\end{array}
\right] ,\;C_{3}=\left[
\begin{array}{l}
123 \\
312
\end{array}
\right] ,\;C_{3}^{2}=\left[
\begin{array}{l}
123 \\
231
\end{array}
\right] \\
\sigma _{v_{1}} &=&\left[
\begin{array}{l}
123 \\
132
\end{array}
\right] ,\;\sigma _{v_{2}}=\left[
\begin{array}{l}
123 \\
321
\end{array}
\right] ,\;\sigma _{v_{3}}=\left[
\begin{array}{l}
123 \\
213
\end{array}
\right]
\end{eqnarray*}
The well known projection operators
\begin{eqnarray*}
\mathcal{P}_{A_{1}} &=&\frac{1}{6}\left( \hat{E}+C_{3}+C_{3}^{2}+\sigma
_{v_{1}}+\sigma _{v_{2}}+\sigma _{v_{3}}\right) \\
\mathcal{P}_{A_{2}} &=&\frac{1}{6}\left( \hat{E}+C_{3}+C_{3}^{2}-\sigma
_{v_{1}}-\sigma _{v_{2}}-\sigma _{v_{3}}\right) \\
\mathcal{P}_{E} &=&\frac{1}{3}\left( 2\hat{E}-C_{3}-C_{3}^{2}\right)
\end{eqnarray*}
will be most useful for present analysis.

One can easily verify that the Hamiltonian
\begin{equation}
H=-\frac{1}{2}\left( \frac{\partial ^{2}}{\partial x_{1}^{2}}+\frac{\partial
^{2}}{\partial x_{2}^{2}}+\frac{\partial ^{2}}{\partial x_{3}^{2}}\right) +%
\frac{1}{2}\left( x_{1}^{2}+x_{2}^{2}+x_{3}^{2}\right) +\xi \left(
x_{1}x_{2}+x_{1}x_{3}+x_{2}x_{3}\right) ,  \label{eq:H_model}
\end{equation}
exhibits $S_{3}$ permutation symmetry and is parity invariant. It describes
a system of three identical particles in a one-dimensional space that
interact with a different one clamped at origin by means of the terms $%
x_{j}^{2}/2$ and between them by means of the terms $\xi x_{i}x_{j}$. It
resembles, for example, the Lithium atom with the nucleus clamped at origin.
One may reasonably argue that this one-dimensional toy model is unsuitable
for the study of atomic systems but the point is that here we are merely
interested in the permutation symmetry of the Hamiltonian operator. The
great advantage of this simple model is that the Schr\"{o}dinger equation is
separable and exactly solvable. It is a simplified version of the oscillator
models widely used by Moshinsky\cite{M96}.

By means of the change of variables\cite{F19}
\begin{equation}
y_{1}=\frac{\sqrt{2}x_{2}}{2}-\frac{\sqrt{2}x_{3}}{2},\;y_{2}=\frac{\sqrt{6}%
\left( 2x_{1}-x_{2}-x_{3}\right) }{6},\;y_{3}=\frac{\sqrt{3}\left(
x_{1}+x_{2}+x_{3}\right) }{3},
\end{equation}
the Hamiltonian becomes
\begin{eqnarray}
H &=&-\frac{1}{2}\left( \frac{\partial ^{2}}{\partial y_{1}^{2}}+\frac{%
\partial ^{2}}{\partial y_{2}^{2}}+\frac{\partial ^{2}}{\partial y_{3}^{2}}%
\right) +\frac{k}{2}\left( y_{1}^{2}+y_{2}^{2}\right) +\frac{k^{\prime }}{2}%
y_{3}^{2},  \nonumber \\
k &=&1-\xi ,\;k^{\prime }=1+2\xi
\end{eqnarray}
We appreciate that there are bound states provided that $-1/2<\xi <1$. Under
this condition the eigenfunctions and eigenvalues are given by
\begin{eqnarray}
&&\psi _{n_{1}n_{2}n_{3}}(x_{1},x_{2},x_{3}) =\phi _{n_{1}}(k,y_{1})\phi
_{n_{2}}(k,y_{2})\phi _{n_{3}}(k^{\prime },y_{3})  \nonumber \\
&&E_{n_{1}n_{2}n_{3}} =\sqrt{k}\left( n_{1}+n_{2}+1\right) +\sqrt{k^{\prime }%
}\left( n_{3}+\frac{1}{2}\right) ,\;  \nonumber \\
&&n_{1},n_{2},n_{3}=0,1,2,\ldots .
\end{eqnarray}
where $\phi (k,q)$ is a normalized eigenfunction of the dimensionless
Hamiltonian $H_{HO}$ for the harmonic oscillator
\begin{eqnarray}
H_{HO}\phi _{n}(k,q) &=&\sqrt{k}\left( n+\frac{1}{2}\right) \phi
_{n}(k,q),\;n=0,1,\ldots  \nonumber \\
H_{HO} &=&-\frac{1}{2}\frac{d^{2}}{dq^{2}}+\frac{k}{2}q^{2}
\end{eqnarray}

Since the symmetric group $S_{3}$ is isomorphic to $C_{3v}$ we can label the
irreps as $A_{1}$, $A_{2}$ (both one-dimensional) and $E$ (two-dimensional).
If we added the inversion, then the suitable group would be $D_{3h}$ (among
others) with irreps $A_{1}^{\prime }$, $A_{2}^{\prime }$, $E^{\prime }$, $%
A_{1}^{\prime \prime }$, $A_{2}^{\prime \prime }$ and $E^{\prime \prime }$,
but we will restrict ourselves to the permutation symmetry. The Hamiltonian (%
\ref{eq:H_model}) exhibits also dynamical symmetry because it commutes with
a set of five operators that depend on the coordinates and conjugate
momenta. Consequently, the degeneracy of the energy levels given by $%
n_{1}+n_{2}+1$ is considerably greater than the one predicted even by $%
D_{3h} $. However, for present purposes it will suffice to consider just $%
C_{3v}$ because we are interested only in the permutation symmetry. Note
that the Hamiltonian operator for Lithium (under the clamped-nucleus
approximation) commutes with the total angular momentum of the electrons $%
L^{2}$ and $L_{z}$; therefore, the symmetric group $S_{3}$ will be
insufficient in this realistic case too. Since the dynamical symmetry is
model-dependent we will omit it from now on.

The variables $y_{1}$ and $y_{2}$ are basis for the irrep $E$ while $y_{3}$
is basis for $A_{1}$. For this reason all the states $\psi _{00j}$ are basis
for $A_{1}$ and the symmetry of the states of the non-relativistic
Hamiltonian is completely determined by the quantum numbers $n_{1}$ and $%
n_{2}$. For example, $\psi _{10j}$ and $\psi _{01j}$ are basis for $E$ and
the three degenerate functions with $n_{1}+n_{2}=2$ are basis for both $%
A_{1} $ and $E$. The state $\psi _{11j}$ is basis for $E$ and by means of
the projection operators we easily verify that the linear combinations $\psi
_{20j}+\psi _{02j}$ and $\psi _{20j}-\psi _{02j}$ are basis for $A_{1}$ and $%
E$, respectively. We can carry out this analysis for every energy
level; for example the four states with $n_{1}+n_{2}=3$ are basis
for $A_{1}$, $A_{2}$ and $A_{3}$. A straightforward calculation
shows that $3\psi _{21j}-\psi _{03j}$, $\psi _{30j}-3\psi _{12j}$
and $\left( \psi _{30j}+\psi _{12j},\psi _{21j}+\psi _{03j}\right)
$ are basis for $A_{1}$, $A_{2}$ and $E$, respectively. In this
case only the basis functions for the irreps $A_{1}$ and $A_{2}$
are eigenfunctions of all the permutation operators.

Let us now turn to the construction of antisymmetric spatial-spin functions.
Since $\mathcal{P}_{A_{2}}=\sqrt{6}\mathcal{A}$ we will resort to this
projection operator for the construction of antisymmetric functions. In the
case of three electrons we expect one quadruplet and two doublets. In order
to determine which non-relativistic functions will appear in a standard CI
calculation we choose an arbitrary function $f(x_{1},x_{2},x_{3})$ and
construct antisymmetric functions according to the following expression
\begin{equation}
\mathcal{P}_{A_{2}}\omega _{i}(x_{1})\omega _{j}(x_{2})\omega _{k}(x_{3})%
\mathcal{P}_{u}f(x_{1},x_{2},x_{3})  \label{eq:space-spin}
\end{equation}
where $u=A_{1},A_{2},E$. The procedure is quite simple: on inserting a
product of three monoelectronic spin states the result may be zero or a
valid Slater determinant. This straightforward calculation shows that the
non-relativistic states that are basis for $A_{1}$ are not allowed by the
principles of quantum mechanics. In other words, the non-degenerate energy
levels $E_{00j}$ will not appear in a CI calculation. The states that are
basis for $A_{2}$ appear in the quadruplet, and those belonging to $E$ in
the doublets. It is worth noting that the conclusions based only on the
permutation of the electron variables are model independent and therefore
apply to more realistic models. For example, in the case of Lithium we
expect approximate antisymmetric spatial-spin functions with spatial parts
that are basis for the irreps $A_{2}$ ($S=3/2$) and $E$ ($S=1/2$). In other
words, we would obtain meaningful results with spin-free basis-set functions
belonging to the symmetry species just mentioned. Also notice that equation (%
\ref{eq:space-spin}) can be easily generalized to any number of electrons
for which we only need the projection operators for the corresponding
symmetric group.

It is not difficult to take into account that the Hamiltonian is
also parity-invariant. We simply apply equation
(\ref{eq:space-spin}) with the projection operators
$\mathcal{P}_{u}$ for the symmetry point group $D_{3h}$
($u=A_{1}^{\prime },A_{2}^{\prime },E^{\prime },A_{1}^{\prime
\prime
},A_{2}^{\prime \prime },E^{\prime \prime }$). The result is that $%
A_{1}^{\prime }$ and $A_{1}^{\prime \prime }$ do not appear in the Slater
determinants, $A_{2}^{\prime }$ and $A_{2}^{\prime \prime }$ appear in the
quadruplet, $E^{\prime }$ and $E^{\prime \prime }$ appear in the doublets.
This result agrees with the analysis of the permutation symmetry of the
hydrogen atoms in $H_{3}^{+}$ carried out, for example, by Bunker and Jensen%
\cite{BJ05} and is called missing levels.

\section{Exactly-solvable four-particle model}

\label{sec:N=4}

The symmetric group $S_{4}$ is isomorphic to $O$ and $T_{d}$ and
we will choose the former point-group symmetry here. In this case
we apply a somewhat different strategy. First, we derive the $24$
matrices that produce
all the permutations of the elements of a four-dimensional column vector $%
\mathbf{x}$. Second, we collect the matrices into their respective group
classes and determine the order (also called period length) of each of them%
\cite{C90}. In this way we derive a one-to-one correspondence between the
matrix classes and those appearing in the character table of the group $O$.
Third, with each matrix $\mathbf{M}_{i}$ we build the corresponding operator
$\hat{M}_{i}$ by means of the well known expression $\hat{M}_{i}f(\mathbf{x}%
)=f\left( \mathbf{M}_{i}^{-1}\mathbf{x}\right) $. In this case we will show
neither the character table nor the projection operators that can be easily
constructed by means of well known expressions\cite{C90}. We will just
discuss the results.

As an illustrative example we resort to the oscillator model
\begin{eqnarray}
H &=&-\frac{1}{2}\left( \frac{\partial ^{2}}{\partial x_{1}^{2}}+\frac{%
\partial ^{2}}{\partial x_{2}^{2}}+\frac{\partial ^{2}}{\partial x_{3}^{2}}+%
\frac{\partial ^{2}}{\partial x_{4}^{2}}\right) +\frac{1}{2}\left(
x_{1}^{2}+x_{2}^{2}+x_{3}^{2}+x_{4}^{2}\right)  \nonumber \\
&&+\xi \left(
x_{1}x_{2}+x_{1}x_{3}+x_{1}x_{4}+x_{2}x_{3}+x_{2}x_{4}+x_{3}x_{4}\right)
\end{eqnarray}
that exhibits the appropriate symmetry. By means of the change of variables
\begin{eqnarray}
y_{1} &=&\frac{1}{\sqrt{2}}\left( x_{1}-x_{4}\right) ,\;y_{2}=\frac{1}{\sqrt{%
2}}\left( x_{2}-x_{3}\right) ,\;y_{3}=\frac{1}{2}\left(
x_{1}-x_{2}-x_{3}+x_{4}\right) ,\;  \nonumber \\
y_{4} &=&\frac{1}{2}\left( x_{1}+x_{2}+x_{3}+x_{4}\right)
\end{eqnarray}
the resulting Hamiltonian is separable an exactly solvable
\begin{eqnarray}
H &=&-\frac{1}{2}\left( \frac{\partial ^{2}}{\partial y_{1}^{2}}+\frac{%
\partial ^{2}}{\partial y_{2}^{2}}+\frac{\partial ^{2}}{\partial y_{3}^{2}}+%
\frac{\partial ^{2}}{\partial y_{4}^{2}}\right) +\frac{1-\xi }{2}\left(
y_{1}^{2}+y_{2}^{2}+y_{3}^{2}\right)  \nonumber \\
&&+\frac{1+3\xi }{2}y_{4}^{2}
\end{eqnarray}
It exhibits bound states when $-1/3<\xi <1$ and its eigenfunctions and
eigenvalues are given by
\begin{eqnarray}
&&\psi _{n_{1}n_{2}n_{3}n_{4}}\left( x_{1},x_{2},x_{3},x_{4}\right) =\phi
_{n_{1}}(k,y_{1})\phi _{n_{2}}(k,y_{2})\phi _{n_{3}}(k,y_{3})\phi
_{n_{4}}(k^{\prime },y_{4})  \nonumber \\
&&k =1-\xi ,\;k^{\prime }=1+3\xi  \nonumber \\
&&E_{n_{1}n_{2}n_{3}n_{4}} =\sqrt{1-\xi }\left( n_{1}+n_{2}+n_{3}+\frac{3}{2}%
\right) +\sqrt{1+3\xi }\left( n_{4}+\frac{1}{2}\right)
\end{eqnarray}
The degeneracy of the states of this oscillator is considerably
greater than the one for the preceding example: $\left(
n_{1}+n_{2}+n_{3}+1\right) \left( n_{1}+n_{2}+n_{3}+2\right) /2$.

The variables $y_{1}$, $y_{2}$, $y_{3}$ are basis for the irrep $T_{2}$ and $%
y_{4}$ is basis for $A_{1}$. For this reason all the states of the form $%
\psi _{000j}$ are basis for $A_{1}$ and the symmetry of the non-relativistic
states is determined by the quantum numbers $n_{1}$, $n_{2}$ and $n_{3}$.
For each value of $n_{4}$ the three degenerate states with $%
n_{1}+n_{2}+n_{3}=1$ are basis for $T_{2}$. The six degenerate states with $%
n_{1}+n_{2}+n_{3}=2$ are basis for $A_{1}$, $E$ and $T_{2}$. The ten
degenerate states with $n_{1}+n_{2}+n_{3}=3$ are basis for $A_{1}$, $E$, $%
T_{1}$ and $T_{2}$. The basis functions for the irrep $A_{2}$
appear in a much higher energy level with $n_{1}+n_{2}+n_{3}=6$.
As in the preceding example only the basis functions for the
irreps $A_{1}$ and $A_{2}$ are eigenfunctions of all the
permutation operators.

In order to determine which non-relativistic spatial functions contribute to
the antisymmetric spatial-spin ones we proceed as in equation (\ref
{eq:space-spin}) adding an additional electron to that expression and
choosing the projection operators for the symmetry point-group $O$. In the
case of four electrons we expect one quintuplet, three triplets and two
singlets. Our results show that the spatial functions that are basis for $%
A_{1}$ and $T_{2}$ are not allowed by the antisymmetry principle.
The basis functions for $A_{2}$, $T_{1}$ and $E$ are responsible
for the quintuplets, triplets and singlets, respectively. This
conclusion is not model dependent because it is based entirely on
the symmetry of the problem and applies, for example, to Beryllium
under the clamped-nucleus approximation.

\section{Further comments and conclusions}

\label{sec:conclusions}

Throughout this paper we have analyzed the connection between the
antisymmetric spatial-spin functions given in terms of Slater
determinants and the eigenfunctions of the non-relativistic
Hamiltonian that are basis for the irreps of the symmetric group
$S_{N}$. We restricted ourselves to the particular cases of $N=3$
and $N=4$ electrons because they can be
analyzed by means of the character tables of the point groups $C_{3v}$ and $%
O $, respectively, that are well known for most chemists and physical
chemists. The exactly solvable models chosen here are suitable for
illustration but are not necessary for obtaining the main theoretical
results. From the point of view of symmetry they are identical to the
Lithium and Beryllium atoms and the particular forms of the Hamiltonians
mimic those atoms in the clamped-nucleus approximation. However, the main
conclusions about the symmetry of the spatial parts of the Slater
determinants also applies to the case of finite nuclear mass. If we remove
the motion of the center of mass and place the coordinate origin at the
nucleus the resulting Hamiltonian exhibits the same symmetry $S_{N}$. The
reason is that the coupling terms that appear in the kinetic-energy operator
(the so called mass polarization terms) do not change the permutational
symmetry of the Hamiltonians\cite{FE10}.

That the main theoretical results derived in this paper are not model
dependent is clearly illustrated by the fact that present analysis of three $%
1/2$-spin identical particles by means of the $D_{3h}$ point group agrees
with the results derived by Bunker and Jensen\cite{BJ05} for the hydrogen
nuclei of the $H_{3}^{+}$ molecule.

\section*{Acknowledgments}
The author is indebted to Diego R. Alcoba and Ofelia O\~na for
their full CI calculations that confirmed present theoretical
results.

\end{document}